\documentstyle[12pt]{article}

\def\bmn{\mbox{\boldmath$\nabla$}}

\def\BB{{\bf B}}
\def\BTB{\tilde{{\bf B}}}
\def\btv{\tilde{{\bf v}}}

\def\BE{{\bf E}}

\def\BTE{\tilde {{\bf E}}}

\def\BM{{\bf M}}
\def\br{{\bf r}}
\def\BR{{\bf R}}

\def\bhr{{\bf \hat r}}
\def\tr{\tilde r}
\def\btr{\tilde{{\bf r}}}

\def\bv{{\bf v}}

\def\dlb{D.Lynden-Bell}

\def\ref{\par \noindent \hangindent=3pc \hangafter=1}

\begin{document}
\centerline{WANDERING AMONG NEWTON WONDERS}

\bigskip

\centerline{\it By D. Lynden-Bell}

\centerline{\it Institute of Astronomy, Cambridge}

\bigskip

\large
In the electromagnetic fields ${\bf B} = (\omega t)^{-2}{\bf b}({\bf
r}/\omega t),\ {\bf E}= -{\bf r} \times {\bf B}/(ct)$ the \linebreak
trajectories of non-relativistic charged particles conserve $({\bf r}
- {\bf v}t)^2$.  The transformation $\tilde{{\bf r}}={\bf r}/(\omega
t)$, $\tau = \omega ^{-2}t^{-1}$ maps such trajectories into orbits in
the constant magnetic field ${\BTB} = - {\bf b}(\tilde{{\bf r}})$ all
of which conserve $\tilde{{\bf v}}^2$. \linebreak$\omega $ is a constant.  The
transformation may also be used to transform any fields obeying
${\rm curl}\ {\bf E} = -c^{-1}\partial {\bf B}/\partial t,\
{\rm div}\ {\bf B} = 0$ into others and relates the particle
trajectories in them.
\normalsize

\bigskip

\noindent{\it Introduction}

Newton's equal areas in equal times theorem for non-coplanar
motions$^{1,2}$ stimulated me to ask under what circumstances could
the {\bf magnitudes} of other vector constants of the motion remain
constant even when extra forces were introduced which change their
direction$^3$.  In particular $({\bf r} - {\bf v}t)^2$ is conserved
when the force is of the form $${\bf F} = (q/c)({\bf v} \times {\bf B}
- {\bf r} \times {\bf B}/t) \eqno (1)$$ for any ${\bf B}({\bf r},\
t)$, which may or may not have zero divergence as in electricity.

The orbits in such a field were derived analytically for the special
case of the electromagnetic field$^3$ $${\bf B} = (\omega t )^{-2}{\bf
b},\ {\bf E} = -{\bf r} \times {\bf B}/(ct) $$ with ${\bf b}$ constant
and $\omega $ a constant of dimension $t^{-1}$ put in to make ${\bf
B}$ and ${\bf b}$ have the same dimensions.

Here we demonstrate an unexpected correspondence between orbits under
any force law of the form (1) and those under the conservative force
law $${\bf F} = (\tilde{{\bf v}}/c) \times \tilde{{\bf B}} \eqno (2)$$ 
where $$\tilde{{\bf r}} = {\bf r}/\theta ,\ \ \theta = \omega t,\
\ \tau =
\omega ^{-1}/\theta ,\ \ \tilde{{\bf v}} = d\tilde{{\bf r}}/d\tau \ ,
\eqno (3)$$ and $$\tilde{{\bf B}}(\tilde{{\bf r}},\ \tau) = -
(\omega \tau )^{-2} {{\bf B}}(\tilde{{\bf r}}/\omega \tau,\ \
\omega^{-2} \tau^{-1}) = -\theta ^2{\bf B}({\bf r},\ t)\ .\eqno (4)$$
The conservation of $({\bf r} - {\bf v}t)^2$ under the force (1)
translates into the conservation of $\tilde{{\bf v}}^2$ under force
(2), because
$$\tilde{{\bf v}} = d
\tilde{{\bf r}}/d\tau = - \omega\theta^2 d({\bf r} /\theta )/d\theta  = {\bf r} -{\bf v}t \ .$$
The equation of motion of a unit mass particle of charge $q$ under the
force law (1) is, $$d^2{\bf r}/dt^2 = (q/c)\left [d{\bf r}/dt \times
{\bf B} -{\bf r} \times {\bf B}/t\right ]\ . \eqno (5)$$ Now under the
transformation (3) equation (5) can be written
$$\omega^{2}\theta^3d^2({\tilde{\bf r}}\theta )/d\theta^2 =
(q/c)\omega ^{2}\theta^2 d \tilde{{\bf r}}/d\theta \times
\theta^2{\bf B}\ . \eqno (6)$$ But $\theta^3d^2({\tilde{\bf r}}\theta )/d\theta^2 = \theta^2 d/d\theta (\theta^2d{\tilde{\bf r}}/d\theta )= \omega^{-2}d^2{\tilde{\bf r}}/d\tau^2$
so we may rewrite (6) in the form,
$$d^2\tilde{{\bf r}}/{\bf d}\tau^2 = (q/c)d\tilde{{\bf r}}/d\tau
\times \tilde{{\bf B}}\ , \eqno (7)$$  where ${\tilde {\bf B}}$ is given by
(4).  Equation (7) is the equation of motion of a unit mass particle
of charge $q$ in a field $\tilde{{\bf B}}$ of magnetic type with
$\tau$ as the time.  So if ${\bf r} = {\bf R}(t)$ is a solution of
(1), $\tilde{{\bf r}} = \omega \tau {\bf R} (\omega^{-2}\tau ^{-1})$
solves (7).

\bigskip

\noindent{\it Application to Electromagnetic orbits}            

Under what circumstances can forces of the form (1) be delivered on a
charge $q$ by electromagnetic fields?  Clearly the first term in the
bracket of (1) is of magnetic type (provided ${\bf B}$ has zero
divergence) and for the other to be electric we need $$\BE=-{\bf r}
\times {\bf B}/(ct)$$ Thus ${\rm Curl}\ {\bf E} = (2{\bf B} + {\bf
r.}\mbox{\boldmath$\nabla$}{\bf B})/(ct)\ .$ \\ Putting this equal to
$-(1/c)\partial {\bf B}/\partial t$ leads to $t
\partial / \partial t (\BB t^2 ) + \br \cdot \bmn (\BB t^2)=0 $ of
which the general solution is$${\bf B} = \theta ^{-2}{\bf b} ({\bf
r}/\theta ) \eqno (8)$$ where ${\bf b}$ is any (vector) function of
its argument subject to Maxwell's condition \linebreak ${\rm div}\
{\bf b} = 0$.  Under the transformation (4) we find that (8) reduces
to ${\bf \tilde{{B}}}= -{\bf b}(\tilde{{\bf r}})$, so $\tilde{{\bf
B}}$ can be any stationary magnetic field.

Thus, if $\btr = \btr (\tau)$ is the trajectory of a charged test
particle that moves \linebreak non-relativistically in any stationary
magnetic field $\BTB = - {\bf b} (\btr)$, then the \linebreak
electromagnetic fields ${\bf B} = \theta^{-2}{\bf b}({\bf r}/\theta),\
\BE=-{\bf r} \times {\bf B}/(ct)$ satisfy Maxwell's \linebreak${\rm curl}\
{\bf E} = -(1/c)\partial \BB/\partial t,\ {\rm div}\ \BB = 0$ and in
them there is a corresponding trajectory $\br = \theta \btr
(\omega^{-1}/\theta)$ which conserves $(\br - \bv t)^2.$

Neither Maxwell's equations, nor the equations of motion involve a
particular zero point for time, so the same arguments hold if we write
to $t-t_0$ wherever we have written $t$ above and the trajectory then
preserves $\left [\br - \bv (t-t_0)\right ]^2$.

A specific example is given by particles trapped by a dipolar field to
form Van Allen belts; the field is $\tilde {\bf B} = (-{\bf M} + 3 \BM \cdot
\bhr \bhr )/\tr^3$ where $\bhr$ is the unit vector $\btr/\tr = \br/r$.
\linebreak Our theorem relates the orbits $\btr = \btr (\tau)$ in this constant
dipole to the orbits \linebreak $\br = \omega (t-t_0)\btr\left ( \omega^{-2}(t-t_0)^{-1}\right )$ in
the dipolar field $\BB$ with moment $-\omega (t-t_0)\BM$  
$$\BB =\omega (t-t_0)(\BM-3\BM\cdot\bhr\bhr)/r^3 \eqno (9)$$ with ${\bf
E}=-\br \times {\bf M}/(cr^3)\ .$ Now in a uniform field the orbits expand
as the field weakens but in this dipolar field we see the orbits shrink
as $t$ approaches $t_0$ and the dipole weakens.  This is because $\btr
= \btr(\tau )$ is confined to the `radiation belts' for all $\tau$ so the
whole of that motion is now shrunk by the initial factor $(t-t_0)$ in
$\br$.  The physics behind this shrinking of the orbit with the dipole
strength lies in the $c{\bf E} \times \BB/B^2$ drift of the orbit.
The ${\bf E}$, which is an inevitable consequence of the changing ${\bf
B}$, is directed toroidally.  For a decreasing dipole the drift is
directed inwards.  It is then of interest to ask whether the
field strength at the guiding centre increases or decreases as the
dipole strength diminishes.  For an equatorial gyration the drift
velocity is ${\bf v}_d = {\bf r}/(t-t_0)$ and $DB^2/Dt = 2\BB \cdot
\left (\partial \BB /\partial t +\bv_d \cdot \bmn B \right ) =
-4B^2/(t-t_0)$ which is clearly positive for $t\leq t_0$ when the
 field is decreasing.  Thus the drift pushes the orbits into regions
 of higher field strength even though the field strength at a given
 position decreases.  This result emphasises the importance of
 considering the electric fields inevitably associated with changing
 magnetic fields.  These electric fields have been included in our
 theorem on changing orbits.  In this particular example the electric
 field is steady and toroidal but its lack of change depends crucially
 on the $r^{-3}$ dependence of the field strength of a dipole and in
 general the $E$ field is time dependent.  Of course (9) is only the
 correct magnetic field of a changing dipole in the `near zone' in
 which its radiation may be neglected.  Notice that ${\rm div}\ {\bf E} = 0 $
 so no charge density is associated with this `pure induction' $\BE$ field.

\bigskip

\noindent{\it Transformations of more general electromagnetic fields}

Although we found the above results by looking at forces that preserve
$(\br -\bv t)^2)$, we have arrived at a strange transformation that
preserves the two Maxwell equations that do not involve sources, so
the transformation can be applied to any electromagnetic field.  Under
the transformation\\

$
\begin{array}{l|l}
\btr = \br / (\omega t) &	   \br = \btr/(\omega\tau) \\

\omega \tau = (\omega t )^{-1} & \omega t = (\omega\tau) ^{-1} \\

\tilde {\BB} (\btr ,\ \tau)
= -(\omega t)^2 \BB(\br , t)
= -(\omega\tau)^{-2}\BB(\btr /\omega\tau ,
\omega^{-2}\tau^{-1}) 		& \BB (\br ,t)= - (\omega\tau)^2 \BTB (\btr ,\ \tau ) \\

\BTE (\btr , \tau )=  (\omega t)^{3} \left [\BE (\br ,\ t) + \br \times \BB
(\br ,\ t)/ct\right ] &
\BE(\br,\ t) =  (\omega\tau)^{3} [\BTE (\btr ,\ \tau) + \\

= (\omega\tau)^{-3} \left [\BE (\btr \omega^{-1}\tau^{-1},\ \omega^{-2}\tau^{-1} ) + \btr
\times \BB (\btr \omega^{-1}\tau^{-1},\ \omega^{-2}\tau^{-1}) /c \right ]& \hspace{1.5cm}+ \btr
\times \BTB (\btr ,\ \tau )/(c\tau) ]
\end{array}
$

\bigskip
\noindent
we write $\partial / \partial \br$ in place of $\bmn$ and deduce

\bigskip
\noindent
$
\begin{array}{lccc}
&\partial /\partial \btr \cdot \BTB = 0 &
\Leftrightarrow &
\partial /\partial \br \cdot \BB = 0 \\

&\partial /\partial\btr\times\BTE +(1/c)\partial\BTB/\partial \tau =0 &
\Leftrightarrow &
\partial /\partial\br \times \BE + (1/c) \partial \BB /\partial t = 0
\\

{\rm and} & & \\

& d^2\btr/d\tau^2 = q \left [\btv/c \times \BTB + \BTE \right ] &
\Leftrightarrow &
d^2\br/dt^2=q(\bv/c \times \BB + \BE ) \\ 

\end{array}
$
\bigskip

Thus if $\BE$ and $\BB$ obey Maxwell's equations with suitable sources
$\rho, {\bf j}$ then $\BTE$ and $\BTB$ will also (with other sources) and
the trajectories of classical non-relativistic particles $\br = \BR
(t)$ map via the transformation into the trajectories $\btr =\omega \tau
\BR(\omega^{-2}\tau^{-1})$ of particles of the same charge/mass ratio under the
fields $\BTE$ and $\BTB$.  We notice that the field of an electric
dipole is invariant while that of a magnetic monopole reverses under
the transformation.  However if we added a time reversal to the
transformation then both the electric dipole and the magnetic monopole
would be invariant.  For a somewhat related theorem on the
gravitational $N$ body problem see$^4$, but note there should be a dot
over $f$ the fourth time it appears there in the mathematics.

\eject

\centerline{\it References}

\bigskip

\noindent (1) I. Newton, {\it Principia, Scholium to Proposition 2}\ (R.Soc.,
    London), 1687.\\

\noindent (2) D. Lynden-Bell \& M. Nouri-Zonoz, Revs of Mod Phys,
    {\bf 70}, 427, 1999.\\

\noindent (3) \dlb, The Observatory,
    (accepted), 2000.\\

\noindent (4) \dlb, The Observatory, {\bf 102}, No 1048, 86, 1982.\\

\end{document}